\documentclass[useamsfonts,mfastrosym]{pasj01}
\begin{document}
\Received{2015/01/30}%{yyyy/mm/dd}
\Accepted{2015/10/02}%{yyyy/mm/dd}

% \submitted{}

\title{THE SUBARU COSMOS 20: SUBARU OPTICAL IMAGING OF THE HST COSMOS
FIELD WITH 20 FILTERS\altaffilmark{*}}
\altaffiltext{*}{Based on data collected at Subaru Telescope, which is
operated by the National Astronomical Observatory of Japan.}

\author{Y.~\textsc{Taniguchi}\altaffilmark{1}}
\altaffiltext{1}{Research Center for Space and Cosmic Evolution, Ehime
University, 2-5 Bunkyo-cho, Matsuyama 790-8577, Japan }

\author{M.~\textsc{Kajisawa}\altaffilmark{1, 2}}
\altaffiltext{2}{Physics Department, Graduate School of Science \&
Engineering, Ehime University, 2-5 Bunkyo-cho, Matsuyama 790-8577,
Japan}

\author{M.~A.~R.~\textsc{Kobayashi}\altaffilmark{1}}
\author{Y.~\textsc{Shioya}\altaffilmark{1}}
\author{T.~\textsc{Nagao}\altaffilmark{1}}
\author{P.~\textsc{Capak}\altaffilmark{3}}
\altaffiltext{3}{California Institute of Technology, MC 105-24, 1200
East California Boulevard, Pasadena, CA 91125, USA}

\author{H.~\textsc{Aussel}\altaffilmark{4}}
\altaffiltext{4}{AIM Unit\'{e} Mixte de Recherche CEA CNRS,
Universit\'{e} Paris VII UMR n158, F-75014 Paris, France}

\author{A.~\textsc{Ichikawa}\altaffilmark{1, 2}}

\author{T.~\textsc{Murayama}\altaffilmark{5}}
\altaffiltext{5}{Astronomical Institute, Graduate School of Science,
Tohoku University, Aramaki, Aoba, Sendai 980-8578, Japan}

\author{N.~\textsc{Scoville}\altaffilmark{3}}
\author{O.~\textsc{Ilbert}\altaffilmark{6}}
\altaffiltext{6}{Aix Marseille Universit\'{e} , CNRS, LAM (Laboratoire
d'Astrophysique de Marseille), UMR 7326, 13388, Marseille, France}

\author{M.~\textsc{Salvato}\altaffilmark{7}}
\altaffiltext{7}{Max Planck Institut f\"ur Plasma Physik and
Excellence Cluster, 85748 Garching, Germany}

\author{D.~B.~\textsc{Sanders}\altaffilmark{8}}
\altaffiltext{8}{Institute for Astronomy, 2680 Woodlawn Drive,
University of Hawaii, Honolulu, HI, 96822, USA}

\author{B.~\textsc{Mobasher}\altaffilmark{9}}
\altaffiltext{9}{Department of Physics and Astronomy, University of
California, Riverside, CA 92521, USA}

\author{S.~\textsc{Miyazaki}\altaffilmark{10}}
\author{Y.~\textsc{Komiyama}\altaffilmark{10}}
\altaffiltext{10}{National Astronomical Observatory of Japan, 2-21-1
Osawa, Mitaka, Tokyo 181-8588, Japan}

\author{O.~\textsc{Le~F\`evre}\altaffilmark{6}}
\author{L.~\textsc{Tasca}\altaffilmark{6}}

\author{S.~\textsc{Lilly}\altaffilmark{11}}
\author{M.~\textsc{Carollo}\altaffilmark{11}}
\altaffiltext{11}{Institute for Astronomy, Department of Physics, ETH
Zurich, CH-8093 Zurich, Switzerland}

\author{A.~\textsc{Renzini}\altaffilmark{12}}
\altaffiltext{12}{INAF - Osservatorio Astronomico di Padova, vicolo
dell'Osservatorio 5, 35122, Padova, Italy}

\author{M.~\textsc{Rich}\altaffilmark{13}}
\altaffiltext{13}{Department of Physics and Astronomy, University of
California, Los Angeles, CA 90095, USA}

\author{E.~\textsc{Schinnerer}\altaffilmark{14}}
\altaffiltext{14}{Max Planck Institut f\"ur Astronomie, K\"onigstuhl
17, Heidelberg, D-69117, Germany}

\author{N.~\textsc{Kaifu}\altaffilmark{15}}
\altaffiltext{15}{The Open University, 2-11, Wakaba, Mihama-ku. Chiba
261-8586, Japan}

\author{H.~\textsc{Karoji}\altaffilmark{16}}
\altaffiltext{16}{Kavli Institute for the Physics and Mathematics of
the Universe (Kavli IPMU, WPI), The University of Tokyo, Chiba
277-8582, Japan}

\author{N.~\textsc{Arimoto}\altaffilmark{17}}
\altaffiltext{17}{Subaru Telescope, 650 N. A'ohoku Place, Hilo, HI
96720, USA}

\author{S.~\textsc{Okamura}\altaffilmark{18}}
\altaffiltext{18}{Department of Advanced Sciences, Faculty of Science
and Engineering, Hosei University, 3-7-2 Kajino-cho, Koganei-shi,
Tokyo 184-8584, Japan}

\author{K.~\textsc{Ohta}\altaffilmark{19}}
\altaffiltext{19}{Department of Astronomy, Graduate School of Science,
Kyoto University, Kitashirakawa, Sakyo-ku, Kyoto 606-8502, Japan}

\author{K.~\textsc{Shimasaku}\altaffilmark{20}}
\altaffiltext{20}{Department of Astronomy, Graduate School of Science,
The University of Tokyo, 7-3-1 Hongo, Bunkyo-ku, Tokyo 113-0033,
Japan}

\author{T.~\textsc{Hayashino}\altaffilmark{21}}
\altaffiltext{21}{Research Center for Neutrino Science, Tohoku
University, Aramaki-Aza-Aoba, Aoba-ku, Sendai 980-8588, Japan}


\maketitle

\begin{abstract}

 We present both the observations and the data reduction procedures of
 the Subaru COSMOS 20 project that is an optical imaging survey of the
 \textit{HST} COSMOS field, carried out by using Suprime-Cam on the
 Subaru Telescope with the following 20 optical filters: 6 broad-band
 ($B$, $g^\prime$, $V$, $r^\prime$, $i^\prime$, and $z^\prime$), 2
 narrow-band (NB711 and NB816), and 12 intermediate-band filters
 (IA427, IA464, IA484, IA505, IA527, IA574, IA624, IA679, IA709,
 IA738, IA767, and IA827\footnote{This intermediate-band filter system
 is dedicated to the Suprime-Cam, which consists of 20
 intermediate-band filters with a spectral resolution of $R = \lambda
 / \Delta \lambda = 20$--26.  An intermediate-band filter would be
 abbreviated as IB.  However, since this intermediate-band filter
 systems is the first such custom filter series, this filter system is
 named as IA by the Subaru Telescope Office.}).  A part of this
 project is described in Taniguchi et al. (2007) and Capak et
 al. (2007) for the six broad-band and one narrow-band (NB816) filter
 data.  In this paper, we present details of the observations and data
 reduction for remaining 13 filters (the 12 IA filters and NB711).  In
 particular, we describe the accuracy of both photometry and
 astrometry in all the filter bands.  We also present optical
 properties of the Suprime-Cam IA filter system in Appendix.

\end{abstract}

%%%%%%%%%%%%%%%%%%%%%%%%%%%%%%%%%%%%%%%%%%%%%%%%%%%%%%%%%%%%%%%%%%%%%%
\section{INTRODUCTION}

\begin{figure}
 %\plotone{f1.eps}
 \includegraphics[width=8cm]{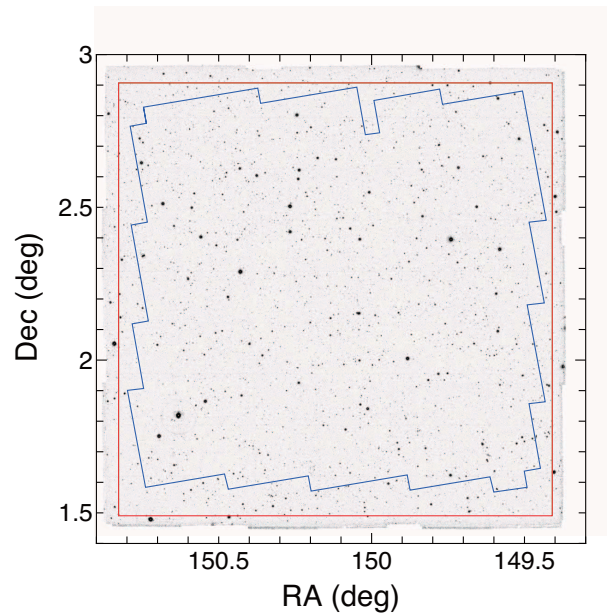}

 \caption{The whole COSMOS field of $1.95~\mathrm{deg}^2$ (red line)
 and the ACS field of $1.64~\mathrm{deg}^2$ (blue line) overlaid on
 the IA427 band image.\label{fig:field}}

\end{figure}

The Cosmic Evolution Survey (COSMOS) is a treasury program on the
\textit{Hubble Space Telescope} (\textit{HST}), awarded a total of 590
\textit{HST} orbits, carried out in Cycles 12 and 13 (Scoville et
al. 2007a, 2007b; Koekemoer et al. 2007).  In total, a sky area of
1.64 square degree is covered with Advanced Camera for Surveys (ACS)
F814W filter around the central position $\mathrm{R.A. (J2000)} =
10^h\ 00^m\ 28.6^s$ and $\mathrm{Decl. (J2000)} = +02^\circ\
12^{\prime}\ 21.0^{\prime\prime}$.  Note that we originally proposed
to map a $1.4~\mathrm{degree} \times 1.4~\mathrm{degree} = 2$~square
degree field.  However, due to the observational constraints, the sky
area of 1.64~square degree was mapped (Koekemoer et al. 2007).  On the
other hand, the Subaru COSMOS 20 project has covered the whole
2~square degree field.  The comparison between the \textit{HST} ACS
field and the Subaru COSMOS 20 field is shown in
Figure~\ref{fig:field}.  A point source limiting magnitude is down to
$\mathrm{AB(F814W)} = 27.2$ ($5\sigma$, 0\farcs 24 diameter aperture).
These ACS observations provide us a large sample of galaxies with a
spatial resolution of 0.1~arcsec covering a redshift range between $z
\sim 0$ to $z \sim 6$ (e.g., Taniguchi et al. 2009; Murata et
al. 2014; Kobayashi et al. 2015).
%----Fig 1

The main purpose of the COSMOS project is to understand the evolution
of galaxies, active galactic nuclei (or super massive black holes),
and dark matter halos together with the evolution of large scale
structures in the Universe.  In order to carry out this project, we
also need multi-wavelength data from X-ray, ultraviolet though optical
to infrared and radio.  Indeed, such multi-wavelength campaign has
been made intensively: X-ray (Hasinger et al. 2007; Elvis et
al. 2009), ultraviolet (Zamojski et al. 2007), optical (Taniguchi et
al. 2007; Capak et al. 2007), infrared (Sanders et al. 2007), and
radio (Schinnerer et al. 2007; Smol{\v c}i{\'c} et al. 2012).  Among
them, optical imaging observations made by use of Suprime-Cam
(Miyazaki et al. 2002) on the Subaru Telescope (Kaifu et al. 2000; Iye
et al. 2004) are highly useful to investigate both photometric
properties and photometric redshifts of the galaxies found in the
COSMOS field (Mobasher et al. 2007; Ilbert et al. 2009; Salvato et
al. 2009).

In our Subaru COSMOS 20 project, we used 20 filters in the optical
covering from 400~nm to 900~nm: six broad-band ($B$, $g^\prime$, $V$,
$r^\prime$, $i^\prime$, and $z^\prime$), twelve intermediate-band
(IA427, IA464, IA484, IA505, IA527, IA574, IA624, IA679, IA709, IA738,
IA767, and IA827), and two narrow-band filters (NB711 and NB816).
This is the origin of the project's name, ``Subaru COSMOS 20''.  Since
we have already given a detailed description on the broad-band and
NB816 imaging of the COSMOS field (Taniguchi et al. 2007; hereafter
Paper~I), we present details of observations, data reductions,
calibration, and quality assessment for the twelve intermediate-band
filters and NB711 in this paper (see also Capak et al. 2007).  As for
the NB711 imaging of the COSMOS field, see also Shioya et al. (2009)
and Kajisawa et al. (2013).  In Appendix, we present a summary of the
optical properties of the Suprime-Cam IA filter system.  Throughout
this paper, we use the AB magnitude system.

The twelve intermediate-band filters are selected from the Suprime-Cam
IA filter set (Hayashino et al. 2000; Taniguchi et al. 2004).  The
spectral resolution of all the IA filters is $R = \lambda / \Delta
\lambda \approx 23$, being just intermediate between typical
broad-band filters ($\lambda / \Delta \lambda \sim 5$) and narrow-band
filters ($\lambda / \Delta \lambda \sim 50$--100).  Therefore, imaging
with multi-IA filters is equivalent to low-resolution spectroscopy
with an $R \approx 23$ (see, for example, Yamada et al. 2005).  It is
also mentioned that the use of IA filters makes it possible to detect
very strong emission-line objects (galaxies or active galactic
nuclei).  Although such objects tend to be rare, some examples have
been discovered to date: ultra strong emission line galaxies (USELs)
defined as $\mathrm{EW (H\beta)} \geq 30$~{\AA} (Kakazu et al. 2007),
and Green Pea objects found in the Galaxy Zoo project (Cardamone et
al. 2009).

Since such very strong emission lines in galaxies affect broad-band
colors of the galaxies (e.g., Nagao et al. 2007), careful analyses are
recommended to make any sample selection of a particular class of
galaxies.  For example, in the case of color selection of very high
redshift galaxies at $z \sim 7$--8, strong emission line galaxies at
$z \sim 2$ with little stellar continuum can act as interlopers (see
Taniguchi et al. 2010; Atek et al. 2011).  On the other hand, such
very strong emission line galaxies themselves are important
populations, because most of them are very metal poor galaxies (Kakazu
et al. 2007; Amor{\'{\i}}n et al. 2014, 2015).  Therefore, surveys of
such objects contribute to the understanding chemical evolution of
galaxies.

In our Subaru COSMOS 20 project, we also use the two narrow-band
filters, NB711 and NB816. Imaging with such narrow-band filters
provides us samples of targeted emission line galaxies.  For example,
NB816 has been used to find Ly$\alpha$ emitters at $z = 5.7$ (e.g., Hu
et al. 2004, 2010; Shimasaku et al. 2005; Murayama et al. 2007).
However, the use of NB816 also provides us samples of H$\alpha$
emitters at $z = 0.24$ (Shioya et al. 2008) and [O~{\sc ii}] emitters
at $z = 1.2$ (Takahashi et al. 2007; Ideue et al. 2009, 2012).  In the
case of COSMOS project, NB711 has also been used to sample both
Ly$\alpha$ emitters at $z = 4.9$ (Shioya et al. 2009) and [O~{\sc ii}]
emitters at $z = 0.9$ (Kajisawa et al. 2013).  Therefore, if we
combine imaging surveys with multiple NB filters, we can trace the
cosmic star formation history from high to low redshifts (e.g.,
Hopkins 2004; Shioya et al. 2008).

Moreover, multi-band optical imaging such as our Subaru COSMOS 20
improves the accuracy of photometric redshifts of galaxies (Ilbert et
al. 2009) and active galactic nuclei (Salvato et al. 2009, 2011).  The
accurate photometric redshifts for the large sample of galaxies allow
us to map the large-scale structure at various redshifts and to study
the environmental effects of the galaxy evolution (Feruglio et
al. 2010; Scoville et al. 2013).  One can also combine such
photometric redshifts with a smaller spectroscopic sample to estimate
the overdensity of galaxies with high accuracy (e.g., Kova{\v c} et
al. 2014) and to measure the clustering strength of AGNs (Georgakakis
et al. 2014).

%%%%%%%%%%%%%%%%%%%%%%%%%%%%%%%%%%%%%%%%%%%%%%%%%%%%%%%%%%%%%%%%%%%%%%
\section{OBSERVATIONS}

\subsection{Observational Strategy}

The COSMOS field covers an area of $1\fdg 4 \times 1\fdg 4$, centered
at $\mathrm{R.A. = 10^h\ 00^m\ 28.6^s}$ and $\mathrm{Decl.} =
+02^\circ\ 12^\prime\ 21.0^{\prime\prime}$.  The Suprime-Cam consists
of ten $2048 \times 4096$~CCD chips and provides a very wide field of
view, $34^\prime \times 27^\prime$ in $10240 \times 8192$~pixels
(0\farcs 202~pixel$^{-1}$) (Miyazaki et al. 2002).  Although the field
of view of the Suprime-Cam had been the widest one among available
imagers on the 8--10~m class telescopes before Hyper Suprime Cam on
the Subaru Telescope (Miyazaki et al. 2012), we needed multiple
pointings to cover the entire COSMOS field.

In our previous Suprime-Cam observations, we used the two dithering
patterns, Pattern~A and Pattern~C (Paper~I).  Pattern~A consists of
$12~\mathrm{shots} \times 4$~sets (48~shots in total) to cover the
whole COSMOS field (see Figure~1 in Paper~I).  This dithering pattern
is a half-array shifted mapping method to obtain accurate astrometry
and a self-consistent photometric solution across the entire field.
Another dithering method is Pattern~C that consists of
$9~\mathrm{shots} \times 4$~sets (36~shots in total) to cover the
whole COSMOS field efficiently (see Figure~2 in Paper~I).  It is noted
that both patterns are designed to take care of spatial gaps
($3^{\prime\prime}$--$4^{\prime\prime}$ or
$16^{\prime\prime}$--$17^{\prime\prime}$) between the CCD chips of the
Suprime-Cam.

After our previous observations, we confirmed that observations with
Pattern~C only are enough to obtain accurate photometry and
astrometry.  The flat frames generated using only Pattern~C in the
broad bands are consistent with those created with both Pattern~A and
Pattern~C within 1\% root mean square.  Therefore, our new
observations with the intermediate and narrow-band filters were made
by using Pattern~C. This made our observations more efficient than our
previous observations.
%----Fig 2
\begin{figure*}
 \begin{center}
  \includegraphics[width=0.7\linewidth]{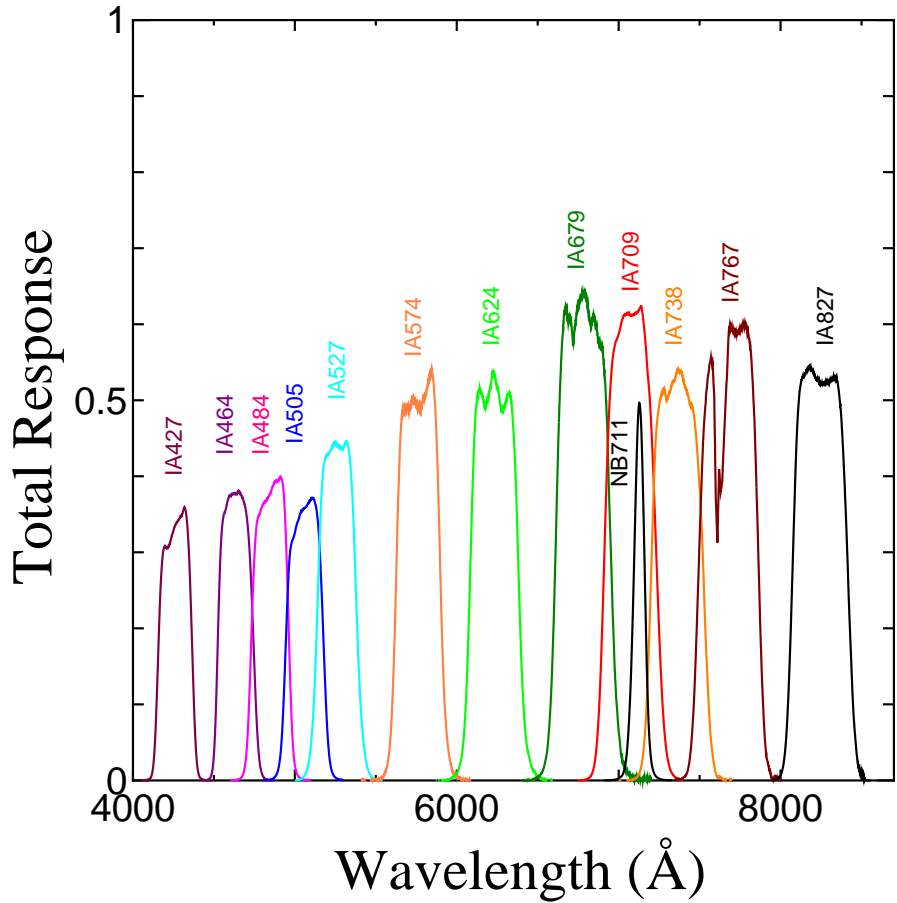}
 \end{center}

 \caption{Filter response curves at the center of the filter,
 including effects of the CCD sensitivity, the atmospheric
 transmission, and the transmission of the telescope and the
 instrument.\label{filter_response}}

\end{figure*}

In this paper, we used twelve intermediate-band filters (IA427, IA464,
IA484, IA505, IA527, IA574, IA624, IA679, IA709, IA738, IA767, and
IA827) and one narrow-band filter (NB711).  Note that we intended to
use another NB filter, NB921, whose effective wavelength and the full
width at half maximum (FWHM) are $\lambda_\mathrm{eff} = 9196$~{\AA}
and $\Delta\lambda = 132$~{\AA}, respectively (Kodaira et al. 2003;
Kashikawa et al. 2004; Taniguchi et al. 2005).  However, we did not
have enough time to take any NB921 data.

\begin{figure*}
 \begin{center}
  \includegraphics[width=0.75\linewidth]{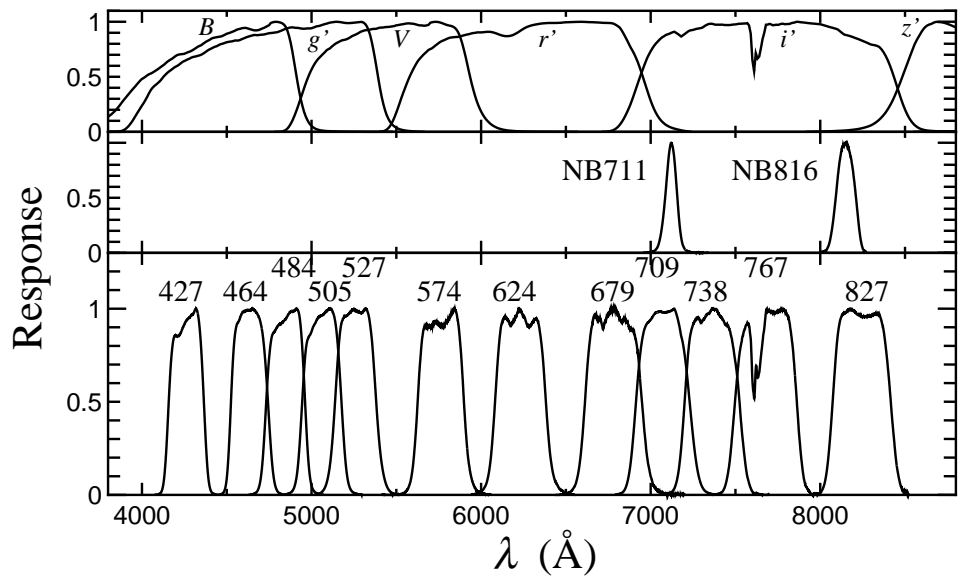}
 \end{center}

 \caption{Filter response curves at the center of the broad-band
 filters (upper panel), narrow-band filters (middle panel), and
 intermediate-band filters (lower panel) used in the Subaru COSMOS 20
 project.  These response curves include the CCD sensitivity, the
 atmospheric transmission, and the transmission of the telescope, the
 instrument, and the filter.  These curves are normalized to a maximum
 response of 1.\label{filter_IAbbNB3}}

\end{figure*}

The filter response curves including the CCD sensitivity and the
atmospheric transmission are shown in Figure~\ref{filter_response} for
the twelve IA filters and NB711 (see also Section 3.2 for details).
In Figure~\ref{filter_IAbbNB3}, we also show those for the 20 filters
used in Subaru COSMOS 20.  In order to see the wavelength coverage
fairly, all the response curves are normalized; i.e., all the peak
values are set to be unity.  Note that the current CCD
chips\footnote{These new CCD chips were installed on July 2008; see
for details the following URL,
http://www.naoj.org/Observing/Instruments/SCam/parameters\_mit.html.}
installed on Suprime-Cam are different from those used in our Subaru
COSMOS 20 project.
%----Fig 3

\subsection{Observational Programs and Runs}

%-----------------------------------------------------
%    Table 1
%-----------------------------------------------------
\begin{table*}[htbp]
 \centering
 \tbl{A summary of observational programs.\label{observational_programs}}{
 \begin{tabular}{clllcc}
  \hline
  Semester & ID No.$^a$ & PI  & Program Title & Nights & Paper$^b$\\
  \hline
  S03B & 239I   & Y. Taniguchi & COSMOS-Broad$^c$  & 10  & 1 \\
  S04A & 080    & Y. Taniguchi & COSMOS-Narrow$^d$ & 2.5 & 1 \\
  S04B & 142I   & Y. Taniguchi & COSMOS-21$^e$     & 8   & 1 \\
  S04B & UH-17A & N. Scoville  & COSMOS-21$^e$     & 4   & 1 \\
  S05B & 013I   & Y. Taniguchi & COSMOS-21$^e$     & 10  & 2 \\
  S06B & 026    & Y. Taniguchi & COSMOS-21$^f$     & 4.5 & 2 \\
  \hline
 \end{tabular}}

 \begin{tabnote}
  $^a$The figure ``I'' given in the last of ID number means an
  Intensive Program.

 $^b$$1 = \mathrm{Paper~I}$, and $2 =
 \mathrm{this~paper}$.

 $^c$Suprime-Cam Imaging of the \textit{HST} COSMOS
 2-Degree ACS Survey Deep Field (Intensive Program).

 $^d$Wide-Field Search for Ly$\alpha$ Emitters at $z =
 5.7$ in the \textit{HST}/COSMOS Field.

 $^e$COSMOS-21: Deep Intermediate \& Narrow-band Survey
 of the COSMOS Field (Intensive Program).

 $^f$COSMOS-21: Deep Intermediate-band Survey of the
 COSMOS Field.
 \end{tabnote}
\end{table*}
%-----------------------------------------------------
%    Table 2
%-----------------------------------------------------
\begin{table*}[htbp]
 \centering
 \tbl{A summary of observational runs.\label{observational_runs}}{
  \begin{tabular}{llcclc}
   \hline
   ID No.$^a$ & Period & Nights  & Avail. Nights & Bands & Paper$^b$\\
   \hline
   S03B-239I & 2004 Jan 16--21 & 6   & 5 & $B$, $r^\prime$, $i^\prime$, $z^\prime$  & 1 \\
   S03B-239I & 2004 Feb 15--18 & 4   & 2 & $V$, $i^\prime$ & 1 \\
   S04A-080  & 2004 Apr 15--19$^c$ & 2.5 & 1 & $NB816$ & 1 \\
   S04B-142I & 2005 Jan 8--10  & 3   & 0 & no data & 1 \\
   UH-17A    & 2005 Feb 3     & 1   & 0 & no data & 1 \\
   S04B-142I & 2005 Feb 9--13  & 5   & 2 & $g^\prime$, $V$, $NB816$ & 1  \\
   UH-17A    & 2005 Mar 10--12 & 3   & 1 & $NB816$ & 1 \\
   S04B-142I & 2005 Apr 1--4$^d$ & 4 & 3 & $g^\prime$, $NB816$ & 1 \\
   S05B-013I & 2006 Jan 27--Feb 1$^e$  & 6   & 6 & $IA427$, $IA574$, $IA709$, $IA827$,  $NB711$ & 2  \\
   S05B-013I & 2006 Feb 22--25  & 4   & 3 & $IA464$, $IA505$, $IA679$ & 2  \\
   S06B-026  & 2006 Dec 17--19$^f$  & 1.5   & 1.5 & $IA624$, $IA679$ & 2 \\
   S06B-026  & 2007 Jan 15--18$^f$  & 2  & 2 & $IA484$, $IA527$, $IA738$ & 2  \\
   S06B-026  & 2007 Mar 20--22$^{e,\ g}$  & 1   & 1 & $IA767$ & 2 \\
   \hline
  \end{tabular}}
  \begin{tabnote}
   $^a$The figure ``I'' given in the last of ID number means an
   Intensive Program.

   $^b$$1 = \mathrm{Paper~I}$, and $2 =
   \mathrm{this~paper}$.

   $^c$First half night was used in every night.

   $^d$Compensation nights because of the poor weather in S04B-142 Jan
   and Feb runs.

   $^e$Observing time exchange with Dave Jewitt (IfA, UH).

   $^f$Second half night was used in every night.

   $^g$Three hours  $\times$  3.
  \end{tabnote}
\end{table*}
Our Suprime-Cam observations of the COSMOS field have been made during
a period between 2006 January and 2007 March, consisting of two
open-use observing programs: S05B-013I (COSMOS-21: Deep Intermediate
\& Narrow-band Survey of the COSMOS Field) and S06B-026 (COSMOS-21:
Deep Intermediate-band Survey of the COSMOS Field).  The first one is
an Intensive Program on the Subaru Telescope.  Fourteen and half
nights were allocated for these two proposals
(Tables~\ref{observational_programs} and \ref{observational_runs}).
With 28.5~nights (including 4 compensation nights) allocated for our
COSMOS observations in Paper~I, 43~nights were allocated in total.

Our observations in the available nights shown in
Table~\ref{observational_runs} were made under the photometric
conditions except for the IA505- and IA679-bands observations on
Feb. 24, 2006.  We observed the spectrophotometric standard stars
immediately before and after the target observations for the
photometric calibration.  These standard stars have been observed at
various airmass, defocussing the telescope to avoid saturation.  The
observed standard stars are summarized in Table~\ref{standard_star}.
Although we also observed GD~71 with several IA filters, all the data
were saturated and we did not use GD~71 for the photometric
calibration.
%-----------------------------------------------------
%    Table 3
%-----------------------------------------------------
\begin{table*}[htbp]
 \tbl{A summary of standard stars.}{
 \begin{tabular}{cl}
  \hline
  Band & Standard stars \\
  \hline
  $B$       & SA~95-193, SA~98-685, SA~101-207, SA~104 \\
  $V$       & SA~95-193, SA~101-207, SA~104 \\
  $g^\prime$ & G~163-51, SA~101-207 \\
  $r^\prime$ & Feige~22, Rubin~149F, SA~95-193, SA~98-685, SA~101-207 \\
  $i^\prime$ & Feige~22, Rubin~149F, Rubin~152, SA~95-193, SA~98-685, SA~101-207 \\
  $z^\prime$ & Feige~22, Rubin~152, SA~95-193, SA~98-685, SA~101-207 \\
  $IA427$   & GD~50, GD~108, HZ~4, HZ~21, HZ~44 \\
  $IA464$   & GD~50, GD~108, HZ~4 \\
  $IA484$   & GD~50, GD~108, HZ~4, HZ~21, HZ~44 \\
  $IA505$   & GD~50, GD~108, HZ~4, HZ~21, HZ~44 \\
  $IA527$   & GD~50, GD~108, HZ~4, HZ~21 \\
  $IA574$   & GD~50, GD~108, HZ~4, HZ~21, HZ~44 \\
  $IA624$   & GD~50, GD~108, HZ~4, HZ~21, HZ~44 \\
  $IA679$   & GD~108, HZ~21, HZ~44 \\
  $IA709$   & GD~50, GD~108, HZ~4, HZ~21, HZ~44 \\
  $IA738$   & GD~50, GD~108, HZ~4, HZ~21 \\
  $IA767$   & GD~50, GD~108, HZ~4, HZ~21, HZ~44 \\
  $IA827$   & GD~50, GD~108, HZ~4, HZ~21, HZ~44 \\
  $NB711$   & GD~50, GD~108, HZ~4, HZ~21, HZ~44 \\
  $NB816$   & Feige~34, GD~50, GD~108, HZ~4\\
  \hline
 \end{tabular}} \label{standard_star}
\end{table*}

%%%%%%%%%%%%%%%%%%%%%%%%%%%%%%%%%%%%%%%%%%%%%%%%%%%%%%%%%%%%%%%%%%%%%%
\section{DATA REDUCTION AND IMAGE QUALITY}

\subsection{Data Reduction}\label{subsec:DataReduction}

All the individual CCD data were reduced using IMCAT\footnote{IMCAT is
distributed by Nick Keiser at
http://www.ifa.hawaii.edu/\~{}kaiser/imcat/} with the same manner as
the Suprime-Cam broad-band data of the COSMOS survey (Capak et
al. 2007).  At first, we performed the bias subtraction and the
masking of bad or saturated pixels.  Then the flat fielding was
carried out with the median dome flats.  We subtracted the median sky
frames from the flat-fielded object frames to remove the night sky
illumination and fringe pattern.  The residual background was measured
in a grid of $128 \times 128$~pixel squares after masking objects and
subtracted.

%After the sky subtraction, we calculated an astrometric solution for
%each CCD chip in all frames using the astrometric reference catalog
%determined from the CFHT $+$ Subaru $i'$-band data. The $i'$-band
%reference catalog was based on the USNO-B1.0 (Monet et al. 2003) and
%the VLA-COSMOS data (Schinnerer et al. 2004; Schinnerer et al.
%2007). The astrometry of the $i'$-band reference data was basically
%registered to the USNO-B1.0 catalog with a large number of matched
%stars, while the VLA-COSMOS data, which have more precise astrometry
%but have a relatively small number of matched objects, were used to
%correct the USNO-B1.0 astrometry.  The internal accuracy of the
%reference catalog is $\Delta \alpha \cos\delta = -26.8 \pm 4.0$ mas
%and $\Delta \delta = -18.6 \pm 4.9$ mas, and the absolute offset to
%the VLA astrometry is $\Delta \alpha \cos\delta = -55.8 \pm 3.8$ mas
%and $\Delta \delta = 81.4 \pm 4.2$ mas.  All images were forced to
%that grid using a 3-5th order polynomial.  The polynomial order was
%increased until the astrometric errors were consistent with the seeing
%size in the data. The internal scatter of the resulting astrometry is
%always less than 0.2~arcsec.
After the sky subtraction, we calculated an astrometric solution for
each CCD chip in all frames using the COSMOS astrometric reference
catalog.  This catalog was build in 2004 using the Megacam
$i^\ast$-band data (Capak et al. 2007), a dataset with bright enough
saturation magnitude on individual exposures ($i^\ast = 16$) to be
registered on classical astrometric references, and deep enough
(reaching $i^\ast = 24$) to allow for the registration of all other
Subaru and ACS COSMOS data.  The COSMOS astrometric reference catalog
was build iteratively via the following four steps: 1) the astrometric
solution for the Megacam images was computed using the Astrometrix
software (Radovich et al. 2001), that was at the time part of the
Terapix pipeline (SCAMP was only introduced in 2005), using the
USNO-B1.0 catalog (Monet et al. 2003) as reference.  2) The
$i^\ast$-band catalog obtained from the Megacam stack was
cross-matched to the COSMOS VLA pre survey (Schinnerer et al. 2004),
and we measured offsets of $\Delta \alpha \cos{\delta} = 16.1 \pm
4.1$~mas and $\Delta \delta = -144.0 \pm 4.0$~mas, without any
indication of variation across the field.  3) The measured offsets
were applied to the input USNO-B1.0 catalog, and the astrometric
solution of the Megacam $i^\ast$-band images was recomputed in the
same manner as step~1.  4) The final COSMOS astrometric reference
catalog was obtained from the $i^\ast$-band Megacam stack produced at
step~3.  The internal accuracy of the COSMOS astrometric reference,
i.e., the maximum value of the residual of the astrometric solution
derived by Astrometrix when fitting the shifted USNO-B1.0 catalog is
$\Delta \alpha \cos{\delta} = -26.8 \pm 4.0$~mas and $\Delta \delta =
-18.6\pm 4.9$~mas, and the absolute offset to the full survey VLA
astrometry is $\Delta \alpha \cos{\delta} = -55.8 \pm 3.8$~mas and
$\Delta \delta = 81.4 \pm 4.2$~mas.  All our COSMOS 20 images were
forced to the COSMOS astrometric reference using a 3--5th order
polynomial.  The polynomial order was increased until the astrometric
errors were consistent with the seeing size in the data.  The internal
scatter of the resulting astrometry is always less than 0.2~arcsec.

We then performed the scattered light correction for the flat as
described in Capak et al. (2007). The dome and sky flat can be
affected by the scattered light at 3--5\% level. The correction factor
for this effect was calculated in each $128 \times 128$~pixel grid so
that the background subtracted fluxes of an object at different
positions of the detector in the different frames have the same
values.  Objects in the all frames were simultaneously used in the
fitting procedure for each band.  In this process, we also added the
additional correction factor for each frame to take account of the
effects of the airmass and non-photometric condition.  These
correction factors for the scattered light in each region and for the
atmospheric condition in each frame were simultaneously determined in
the fitting for each band (equations (1) and (2) in Capak et
al. 2007).  Thus we made the corrected flat frames and applied them to
the object frames.

Then the frame to frame offsets of the background-subtracted fluxes of
objects were examined as a function of airmass and Modified Julian
Date.  If the data followed the airmass trend estimated from the
standard star observations within the expected 1--2\% error due to
point spread function (PSF) variation, the data were deemed
photometric.  If data stopped following the expected trend, or did not
follow it for a night, the data were deemed non photometric.  The
non-photometric frames were scaled to the mean of the airmass
corrected photometric data.  The frames with extinction greater than
0.5~mag were discarded.  In the case of the IA679 band, where no
photometric data were obtained, all the object frames were scaled to
the least extinct frame.  Therefore the photometry in the IA679 band
should be used with caution.

The flux-matched frames were then smoothed to the same PSF FWHM using
a Gaussian kernel.  After the resampling onto the final astrometric
grid, the PSF-matched frames were combined with a weight of the
inverse variance of each frame, clipping outlier values at more than
$5\sigma$ from the median value in the calculation of each pixel.  In
this procedure, we also generated a root mean square (rms) map that
reflects the true pixel-to-pixel rms, which is the value expected if
the effects of the resampling and smoothing do not exist.  The rms
measured in a given area on this rms map represents the variance that
would be measured in the background of the same area if the variance
was calculated on the individual images that went into the final
mosaiced image.  The PSF sizes of final images are summarized in
Table~\ref{optical_imaging}.
%Note that the PSF size of the images
%used for generating the official photometric catalog was matched to
%that for the image with worst seeing (CTIO/KPNO $K_{\rm s}$-band
%image; see, Capak et al. 2007).  
In addition to these PSF-matched combined images, we also provided the
original-PSF images for each band from the frames that were not
convolved in order to provide a maximum sensitivity for detection of
(compact) sources.  Note that the PSF varies as a function of position
in these original-PSF images, and therefore the color measurements
with a relatively small aperture can be less reliable.  These reduced
images were divided into tiles with a dimension of $10^\prime \times
10^\prime$ as shown in Figure~5 of Paper~I.

For the color measurements and generating the official multi-band
photometric catalog, we additionally convolved the PSF-matched
combined images to match the PSF among all the optical--NIR data from
$U$ to $K_{\rm s}$ band. These data were convolved with a Gaussian
kernel so that the flux ratio between a $3^{\prime\prime}$ and
$10^{\prime\prime}$ aperture for a point source in each band is the
same as that of CTIO/KPNO $K_{\rm s}$-band data, which have the lowest
flux ratio of $\sim 0.75$ (Capak et al. 2007).  The width
($\sigma$-value) of the Gaussian kernel used in the convolution of the
IA and $NB711$-band data is shown in the last column of Table~4.
Using these convolved data, we carried out the multi-band photometry
with a $3^{\prime\prime}$ diameter aperture and the results are
presented in the official photometric catalog.  Note that the matching
of the flux ratio between $3^{\prime\prime}$ and $10^{\prime\prime}$
apertures for point sources does not necessarily guarantee the same
flux ratio for extended sources, because the detailed shapes of the
PSF are not the same among the different bands.  If one needs to
measure colors for extended sources with high accuracy, the
photometric values in the official catalog should be used with
caution.

\subsection{Photometric Calibration}

%----Fig 4
\begin{figure*}
 \begin{center}
 \includegraphics[width=0.8\linewidth]{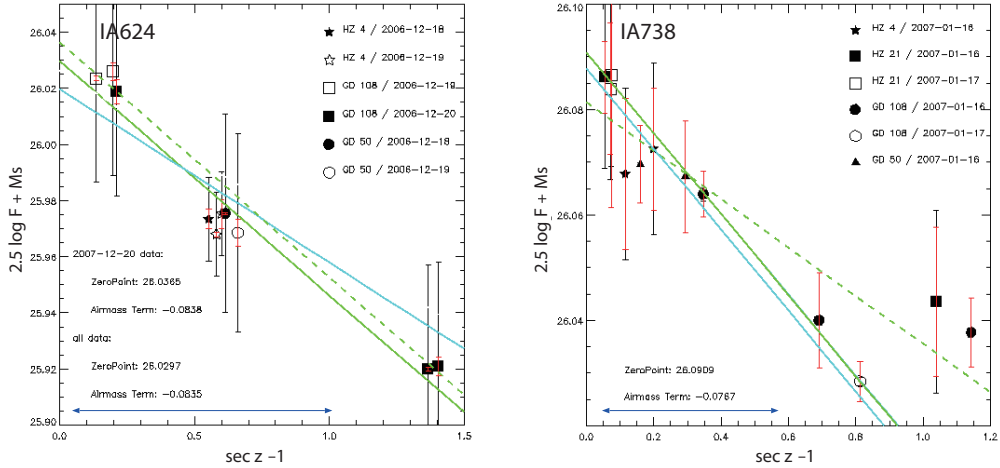}
 \end{center}

 \caption{Determination of the photometric zero points for the IA624
 (left) and IA738 (right) bands. Each symbol represents a different
 standard star and a different night of observation.  The red error
 bars correspond to the uncertainty of the theoretical magnitudes (see
 Table~\ref{tab:standards}) and the black ones represent the combined
 counts measurements and uncertainty of the standard.  The airmass
 relation obtained from the object frames of the survey is given by
 the cyan line.  The green dashed line is a fit to all data of the
 standard stars, while the solid green line is a fit excluding nights
 where there was indication that the standard measurements were not
 obtained under photometric conditions (i.e., the data represented as
 the filled symbols).  The agreement between the two airmass
 determination is reasonable for IA624, and good for IA738 for the
 standard data obtained on 2007-01-17 (the open symbols), while quite
 poor for the standard data obtained in 2007-01-16 (the filled
 symbols).  The range of $(\sec{z} - 1)$ for the targets is shown by
 the blue horizontal arrow in each panel. \label{fig:standards}}

\end{figure*}
Since we used the original intermediate- and narrow-band filters in
this project, the theoretical synthetic magnitudes of the
spectrophotometric standard stars are necessary for the photometric
calibration.  The theoretical complete system response in each band
was computed by multiplying the filter transmission, the Subaru
telescope mirror reflectivity, the prime focus unit's transmission,
the CCD quantum efficiency, and the atmosphere transmission for an
airmass of 1.2.  For the atmosphere, we used the same model as the one
used for the broad-band Suprime-Cam filters (K. Shimasaku, private
communication\footnote{http://hikari.astron.s.u-tokyo.ac.jp/work/suprime/filters/}).
We computed the theoretical synthetic magnitude in the AB system of
our standard stars using the available {\tt CALSPEC\footnote{See
http://www.stsci.edu/hst/observatory/crds/calspec.html.}} spectra,
i.e., {\tt gd50\_004.fits} for GD~50, {\tt gd108\_005.fits} for
GD~108, {\tt hz4\_stis\_001.fits} for HZ~4, {\tt hz21\_stis\_001.fits}
for HZ~21 and {\tt hz44\_stis\_001.fits} for HZ~44 (Bohlin 1996;
Bohlin et al. 2001).  We present these theoretical magnitudes in the
Appendix~\ref{sec:app-B}.  Note that for the HZ stars, the {\tt
CALSPEC} spectra have been interpolated between 5150~{\AA} and
5212~{\AA}, resulting in a larger overall uncertainty in the
calibration of the IA505 and IA527 bands, because this more uncertain
region lies on the edges of these two passbands.

These standard stars have been observed at various airmass as
mentioned above, and we determined the zero point for each band taking
account of the airmass dependence.  If $F$ are the counts measured
from the standard star under an airmass $\sec{z}$, the relation
between the standard magnitude $M_\mathrm{s}$ and the zero point
$Z_\mathrm{p}$ is:
\begin{equation}
-2.5 \log F + Z_\mathrm{p} - M_\mathrm{s} = k (\sec{z} - 1)
\end{equation}
So that the airmass term $k$ is the slope of the expected linear fit
to the data of the standard stars and its intercept gives the zero
point.  Figure~\ref{fig:standards} shows typical examples of the
calibration for the IA624 and the IA738 bands. For most bands, our
zero point determination is accurate within 0.02~mag. In a few cases,
there was a lack of standard star observations at high airmass that
prevented us to derive the airmass term from the standard star data.
In this case, we used the airmass point of the object data for the
zero-point determination.  Note that for the IA679 band, we obtained
no photometric data both for the COSMOS field and the standard stars,
and we scaled to the photometry of the neighboring bands (i.e., IA624
and IA709) assuming the objects were flat in $F_{\nu}$ for the
interpolation.  After the photometric calibration, all the reduced
images are converted to be in units of nanojanskys per pixel (the zero
point of 31.4~mag in the AB magnitude system).

We also checked the consistency among the zero points in the different
bands through the spectral energy distribution fitting for a large
number of galaxies with spectroscopic redshift (Ilbert et al. 2009).
The multi-band photometry from UV to MIR wavelength including the IA-
and narrow-band data were fitted with population synthesis models, and
the systematic difference between the model and observed magnitudes in
a certain band is considered to reflect the zero-point offset.  The
details of the method to determine the zero-point offsets are
described in Ilbert et al. (2006, 2009).  These offsets of the
photometric zero points are shown in the second-last column of
Table~\ref{optical_imaging}.  We note that the offset for the IA679
band is much higher than the other IA bands, which probably reflect
the larger uncertainty in the photometric calibration for this band
mentioned above.  The zero-point offsets in
Table~\ref{optical_imaging} are calculated for the upgraded version
(v2.0) of the photometric redshift catalog from Ilbert et al. (2009)
including the new UltraVISTA data from the DR1 (McCracken et
al. 2012).  Therefore, the offsets in Table~\ref{optical_imaging} are
slightly different from those in Table~1 of Ilbert et al. (2009).
%-----------------------------------------------------
%    Table 4
%-----------------------------------------------------
\begin{table*}
 \tbl{A summary of the optical imaging data for COSMOS.\label{optical_imaging}}{
 \begin{tabular}{cccccccrc}
  \hline
  Band & $\lambda_\mathrm{eff}^a$ & FWHM$^b$ &  TDT$^c$ & $m_{\rm lim}^d$ & $\sigma_{m_{\rm lim}}^e$ & PSF FWHM$^f$ & offset$^g$ & $\sigma_{\rm conv}^h$\\
  & (\AA) & (\AA) & (min) & (mag)  & (mag) & ($^{\prime\prime}$) & (mag) & ($^{\prime\prime}$)\\
  \hline
  $IA427$ & 4263.5 & 207.3 & 41.3 & 25.8 & 0.12 & 1.64 & 0.042 & 0.30 \\
  $IA464$ & 4635.1 & 218.1 & 40.0 & 25.6 & 0.13 & 1.89 & 0.040 & 0.29 \\
  $IA484$ & 4849.2 & 229.1 & 36.7 & 25.9 & 0.16 & 1.14 & 0.014 & 0.59 \\
  $IA505$ & 5062.5 & 231.5 & 36.0 & 25.6 & 0.13 & 1.44 & 0.013 & 0.48 \\
  $IA527$ & 5261.1 & 242.7 & 36.7 & 25.7 & 0.12 & 1.60 & 0.041 & 0.41 \\
  $IA574$ & 5764.8 & 272.8 & 45.3 & 25.4 & 0.10 & 1.71 & 0.085 & 0.11 \\
  $IA624$ & 6232.9 & 299.9 & 36.7 & 25.7 & 0.16 & 1.05 & 0.009 & 0.66 \\
  $IA679$ & 6781.1 & 335.9 & 41.3 & 25.3 & 0.10 & 1.58 & $-0.176$ & 0.31 \\
  $IA709$ & 7073.6 & 316.3 & 40.0 & 25.4 & 0.10 & 1.58 & $-0.021$ & 0.14 \\
  $IA738$ & 7361.5 & 323.8 & 37.0 & 25.4 & 0.10 & 1.08 & 0.021 & 0.59 \\
  $IA767$ & 7684.9 & 365.0 & 45.0 & 25.1 & 0.13 & 1.65 & 0.039 & 0.20 \\
  $IA827$ & 8244.5 & 342.8 & 72.0 & 25.1 & 0.15 & 1.74 & $-0.013$ & 0.13 \\
  $NB711$ & 7121.7 &  72.5 & 35.0 & 25.0 & 0.13 & 0.79 & 0.014 & 0.72 \\
  \hline
  \end{tabular}}
 \begin{tabnote}

  $^a$Effective wavelength calculated from the filter response curve
  including the effects of the CCD sensitivity, the atmospheric
  transmission, and the transmission of the telescope and the
  instrument shown in Figure~\ref{filter_response}.

  $^b$FWHM of the filter response curve mentioned above.

  $^c$The target dedicated time.

  $^d$The average 3$\sigma$ limiting magnitude in the AB system within
  $3^{\prime\prime}$ diameter aperture.

  $^e$The standard deviation of $m_\mathrm{lim}$ measured in the 81
  tiles.

  $^f$The PSF size of the final images. Note that the PSF of each
  filter band is finally matched so that the flux ratio between a
  $3^{\prime\prime}$ and $10^{\prime\prime}$ apertures is the same as
  that in the CTIO/KPNO $K_{\rm s}$-band data to provide official
  photometric catalog (see text).

  $^g$Systematic offset of the photometric zero point for each filter
  (see text).

  $^h$The $\sigma$-value of the Gaussian kernel used for the PSF
  matching among the different bands (see text in
  Section~\ref{subsec:DataReduction}).

 \end{tabnote}

\end{table*}

Note that the magnitudes in the public catalog are not corrected for
the Galactic extinction.  Instead, we provide the Galactic extinction
value, $E(B-V)$, from Schlegel et al. (1998) for each object in the
catalog.  The correction for the Galactic extinction in each band can
be calculated from these values.

\subsection{Data Quality}

We estimated the limiting magnitudes using the 81 tiles (the COSMOS
\textit{HST}/ACS field) for each band.  For each tile, we set 50,000
random points and performed aperture photometry with a
$3^{\prime\prime}$ diameter aperture on the PSF-matched images which
were convolved to the resolution of the COSMOS $K_{\rm s}$-band image.
In order to measure the background fluctuation properly, we masked
objects on the images. We used SExtractor version 2.3.2 (Bertin \&
Arnouts 1996) with the detection criteria of 5-pix connection above
the $2\sigma$ significance. Then we replaced the masked regions with
pseudo noise images, which were provided from randomly-shifted
object-masked images.  Then we evaluated the limiting magnitudes from
the standard deviation for the distribution of the random photometry.

\begin{figure*}
 \begin{center}
  \includegraphics[width=0.9\linewidth]{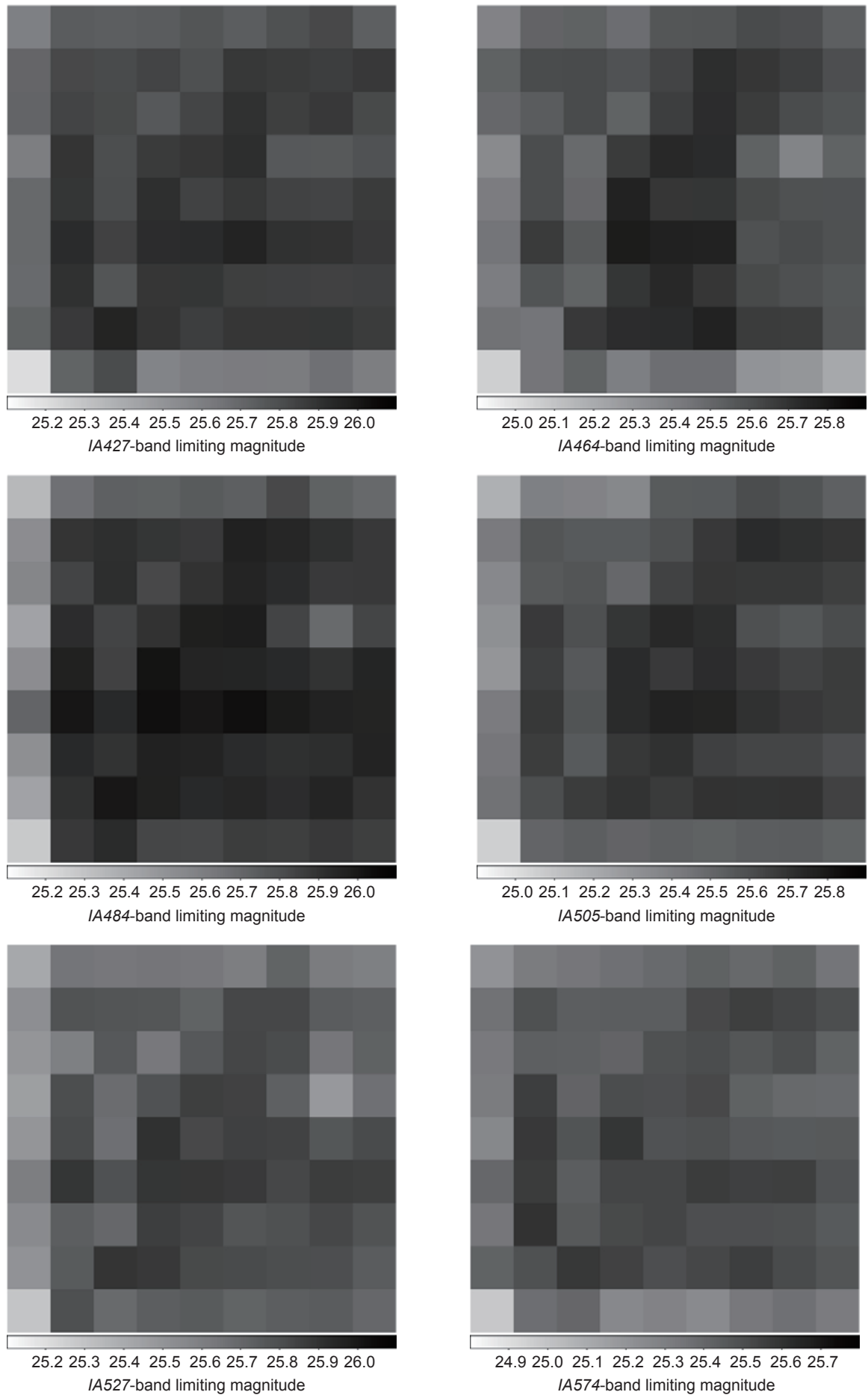}
 \end{center}

 \caption{Variations of $3\sigma$ limiting magnitudes within
 $3^{\prime \prime}$ diameter aperture in the 81 tiles of the
 $IA427$--, $IA464$--, $IA484$--, $IA505$--, $IA527$--, and
 $IA574$--band data from top-left to
 bottom-right.\label{fig:limitmag_IA1}}

\end{figure*}

\begin{figure*}
 \begin{center}
  \includegraphics[width=0.9\linewidth]{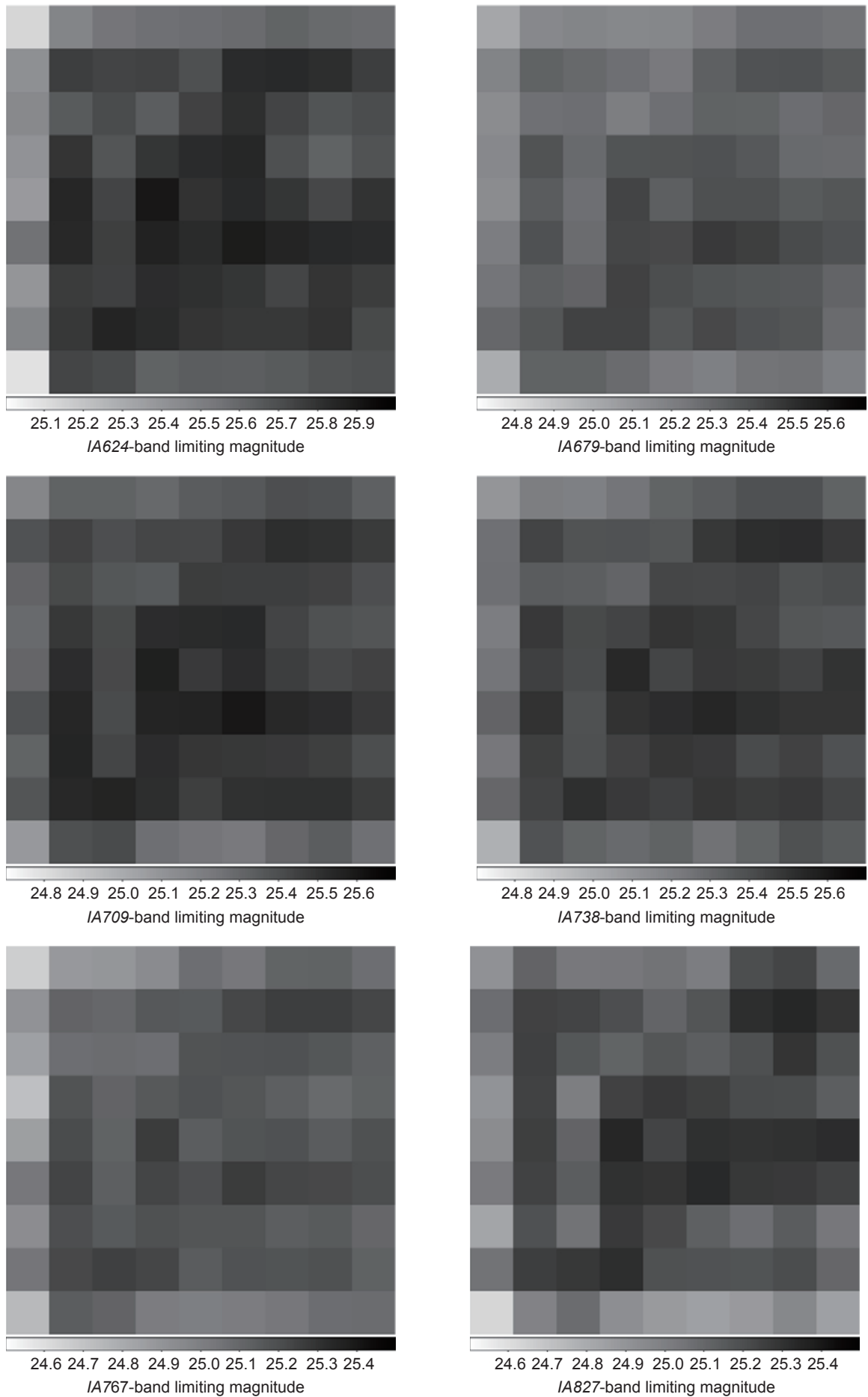}
 \end{center}

 \caption{Same as Figure~\ref{fig:limitmag_IA1}, but for $IA624$--,
 $IA679$--, $IA709$--, $IA738$--, $IA767$--, and $IA827$--band data
 from top-left to bottom-right.\label{fig:limitmag_IA2}}

\end{figure*}

\begin{figure}
 \begin{center}
  \includegraphics[width=7.cm]{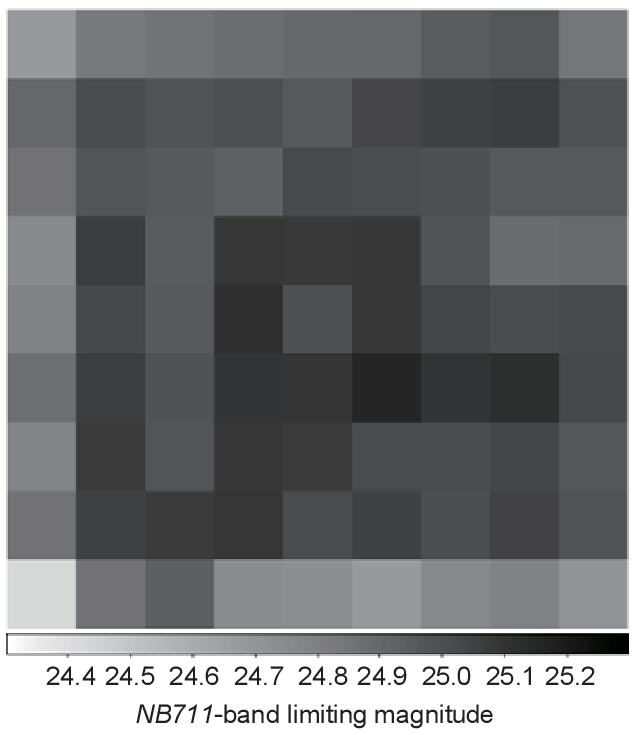}
 \end{center}

 \caption{Same as Figure~\ref{fig:limitmag_IA1}, but for $NB711$--band
 data.\label{fig:limitmag_NB711}}

\end{figure}

The average limiting magnitudes of the 81~tiles for the IA and NB711
bands are listed in Table~\ref{optical_imaging}.  As shown in
Table~\ref{optical_imaging} and Figures~\ref{fig:limitmag_IA1} and
\ref{fig:limitmag_IA2}, the $3\sigma$ limiting magnitudes are $\sim
25.1$--25.9~mag in IA427--IA827 bands.  The NB711 data reach to the
limiting magnitude of $\sim 25.0$~mag as shown in
Table~\ref{optical_imaging} and Figure~\ref{fig:limitmag_NB711}.  The
standard deviation of the limiting magnitudes among the 81 tiles for
each band is $\sim 0.10$--0.16~mag.  As seen in
Figures~\ref{fig:limitmag_IA1}--\ref{fig:limitmag_NB711}, the limiting
magnitudes are brighter in the tiles at the edge of our survey field,
because the total exposure time is smaller in these regions.  Some
tiles where very bright stars illuminate surrounding sky region also
show brighter limiting magnitudes.
%----Fig 5
%----Fig 6
%----Fig 7

%%%%%%%%%%%%%%%%%%%%%%%%%%%%%%%%%%%%%%%%%%%%%%%%%%%%%%%%%%%%%%%%%%%%%%
\section{DISCUSSION}

We present deep optical imaging observations made with the Suprime-Cam
on the Subaru Telescope with 20 filters [6 broad-band, 12
intermediate-band (IA), and 2 narrow-band (NB) filters]: Subaru COSMOS
20.  In this paper, we describe the details of our imaging with the 12
IA filters and NB711.  Note that those of the other seven filters are
given in Paper~I.

The use of intermediate-band filters has generally a couple of
scientific merits: (1) improvement of the accuracy of photometric
redshifts and (2) selection of very strong emitters.  First, we
discuss the improvement of the accuracy of photometric redshifts.  As
described in Mobasher et al. (2007), our previous accuracy of
photometric redshifts based on six Subaru broad band, CFHT $u$ band,
ACS F814W, NB816, and CTIO/KPNO $K_{\rm s}$ photometric data is
$\sigma_{\Delta z} = 0.031$ where $\Delta z = (z_{\rm phot} - z_{\rm
spec}) / (1 + z_{\rm spec})$; note that $z_{\rm phot}$ and $z_{\rm
spec}$ are photometric and spectroscopic redshifts, respectively.
Since both $u$ and $K_{\rm s}$ are used together with optical data,
the accuracy of $z_{\rm phot}$ is better than that of typical optical
studies (e.g., Hogg et al. 1998).  However, in the COSMOS project,
thanks to its multi-wavelength campaign, 30 band photometric data
including Subaru COSMOS 20 data are accumulated to obtain much more
accurate estimates of $z_{\rm phot}$ (Ilbert et al. 2009; see also
Salvato et al. 2009, 2011).  The accuracy of $z_{\rm phot}$ is
improved to $\sigma_{\Delta z} = 0.007$ for $i < 22.5$.  Even at
fainter magnitudes of $i < 24$, the accuracy is found to be still high
as $\sigma_{\Delta z} = 0.012$ for the galaxies at $z_\mathrm{spec} <
1.25$ (Ilbert et al. 2009).

There are several similar surveys with the use of intermediate band
filters.

[1] COMBO-17 (Classifying Objects by Medium-Band Observations in 17
Filters): This is a pioneering optical survey with multi-band filters
(Wolf et al. 2003). The COMBO-17 covers three $30^{\prime} \times
30^{\prime}$ fields, including the Extended Chandra Deep Field South
(ECDF-S).  In this survey, twelve intermediate-band filters were used
together with five broad-band ones ($U$, $B$, $V$, $R$, and $I$) by
using the Wide Field Imager at the MPG/ESO 2.2~m telescope on La
Silla, Chile.  Their intermediate-band filters cover 410~nm to 920~nm.
The spectral resolution is not fixed for all the filters but ranges
from $R = \lambda / \Delta \lambda = 14$ to 61 (mostly from 30 to 40).
The use of intermediate-band filters improves the accuracy of
photometric redshifts to $\sigma_{\Delta z} = 0.007$ for $R < 24$
(Wolf et al. 2004).  This enables them to construct a large sample of
AGNs at $z \sim 1$--5.

[2] MUSYC (the Multiwavelength Survey by Yale-Chile): In this project,
18 intermediate-band filters in the IA filter system for Suprime-Cam
on the Subaru Telescope were used together with 14 board-band data
from optical to mid-infrared (seven optical filters from $U$ to $z$,
three near-infrared filters, $J$, $H$, and $K$, and four Spitzer IRAC
bands, 3.6, 4.5, 5.8, and 8.0~$\mu$m) (Cardamone et al. 2010).  These
data cover a $30^{\prime} \times 30^{\prime}$ field of the ECDF-S,
which is one of the MUSYC fields.  The use of IA filters improves the
accuracy of photometric redshifts at $z = 0.1$ to 1.2 and $z \geq
3.7$: $\sigma_{\Delta z} = 0.01$ for $R < 25$, see Table~8 in
Cardamone et al. (2010) in more detail.  This is attributed that the
Balmer break (3648~{\AA}) or Lyman break (912~{\AA}) falls in
wavelength interval covered by the 18 IA filters.  According to
Cardamone et al. (2010), {\it the use of IA filters not only tightens
the accuracy of photometric redshifts but also can help to rule out
false redshift solution (so called catastrophic failures)}.

[3] MAHOROBA-11: This survey is a scaled down version of Subaru COSMOS
20 (Yamada et al. 2005).  In this survey seven IA filters are used
together with five broad-band filters.  These data cover a
$34^{\prime} \times 27^{\prime}$ area in the Subaru XMM-Newton Deep
Survey field.  Their main purpose is to search for Ly$\alpha$ emitters
at $z > 3.7$ by using a photometric redshift method.  They showed that
the fraction of false detection is only 10\%.

[4] ALHAMBRA (the Advanced Large Homogeneous Area Medium-Band Redshift
Astronomilca): This survey has been carried out by using the
wide-field optical camera, Large Area Imager for Calar Alto (LAICA) on
the Calar Alto 3.5~m telescope with 20 intermediate-band filters with
300~{\AA} spacing (Moles et al. 2008; Molino et al. 2014).  The
surveyed area size is 2.79~deg$^2$.  The accuracy of photometric
redshifts is $\sigma_{\Delta z} = 0.01$ for $I < 22.5$ and
$\sigma_{\Delta z} = 0.014$ for $22.5 < I < 24.5$.

In this way, a number of optical wide-field deep surveys have been
carried out by using their original intermediate-band filter systems.
The main reason for this is to obtain more reliable photometric
redshifts for large numbers of objects in the individual surveys; see
Figure~1B in Molino et al. (2014) for a comprehensive comparison among
available optical surveys including surveys with broad-band filters
only such as HDF, SDSS, and so on.  The Subaru COSMOS 20 is the widest
survey among the deep ($m_\mathrm{lim} \sim 25$) optical
intermediate-band surveys.  Some efficient multiple-object
spectrographs are available on 8~m class telescopes (e.g., VIMOS on
the VLTs and FMOS on the Subaru Telescope).  However, imaging surveys
with intermediate-band filters are more efficient to obtain redshift
information for large numbers of objects.

We mention about our future works on study of strong emission-line
objects.  The wide imaging with the IA filter set of the Subaru COSMOS
20 enables us to detect very strong emission-line objects (star
forming galaxies and AGNs) over a extremely large volume.  In our
forthcoming papers, we will present a large sample of IA-excess strong
emission-line objects (Kajisawa et al. 2015, in preparation) and a new
population of MAESTLO ($=$ MAssive Extremely STrong Ly$\alpha$
Emitters) at $z\sim 3$ with rest-frame Ly$\alpha$ equivalent width of
$\rm EW_0({\rm Ly}\alpha) \geq 100$~{\AA} and $M_{\rm star} \geq
10^{10.5}~M_\odot$ (Taniguchi et al. 2015).

Finally, we note that the major COSMOS datasets including the Subaru
images and catalogs are publicly available (following calibration and
validation) through the web site for IPAC/IRSA:\\ {\bf
http://irsa.ipac.caltech.edu/data/COSMOS/}.

%-------------------------------------------------------------------------
\section*{ACKNOWLEDGMENTS}
The \textit{HST} COSMOS Treasury program was supported through NASA
grant HST-GO-09822.  We gratefully acknowledge the contributions of
the entire COSMOS collaboration consisting of more than 70 scientists.
More information on the COSMOS survey is available at {\bf
http://www.astro.caltech.edu/\~{}cosmos}.  It is a pleasure the
acknowledge the excellent services provided by the NASA IPAC/IRSA
staff (Anastasia Laity, Anastasia Alexov, Bruce Berriman and John
Good) in providing online archive and server capabilities for the
COSMOS datasets.  We are deeply grateful to the referee for his/her
useful comments and excellent refereeing, which helped us to improve
this paper very much.  We would also like to thank the staff at the
Subaru Telescope for their invaluable help.  In particular, we would
like to thank Hisanori Furusawa because his professional help as a
support scientist made our Suprime-Cam observations successful.  Data
analysis were in part carried out on common use data analysis computer
system at the Astronomy Data Center, ADC, of the National Astronomical
Observatory of Japan.  This work was financially supported in part by
JSPS (YT: 15340059, 17253001, 19340046, 23244031, TN: 23654068 and
25707010) and by the Yamada Science Foundation (TN).

%-------------------------------------------------------------------------

\appendix

\renewcommand{\thesection}{A}
 \section{Intermediate-band filter system for Suprime-Cam}

 In this section, we present optical properties of the IA filter
 system for Suprime-Cam on the Subaru Telescope.  This filter system
 was developed as a private type of filters by the two authors (TH and
 YT).  Early short descriptions on this filter system are given in
 Hayashino et al. (2000) and Taniguchi (2004).

\begin{figure*}
 \renewcommand{\thefigure}{A1}
 \begin{center}
  \includegraphics[width=\linewidth]{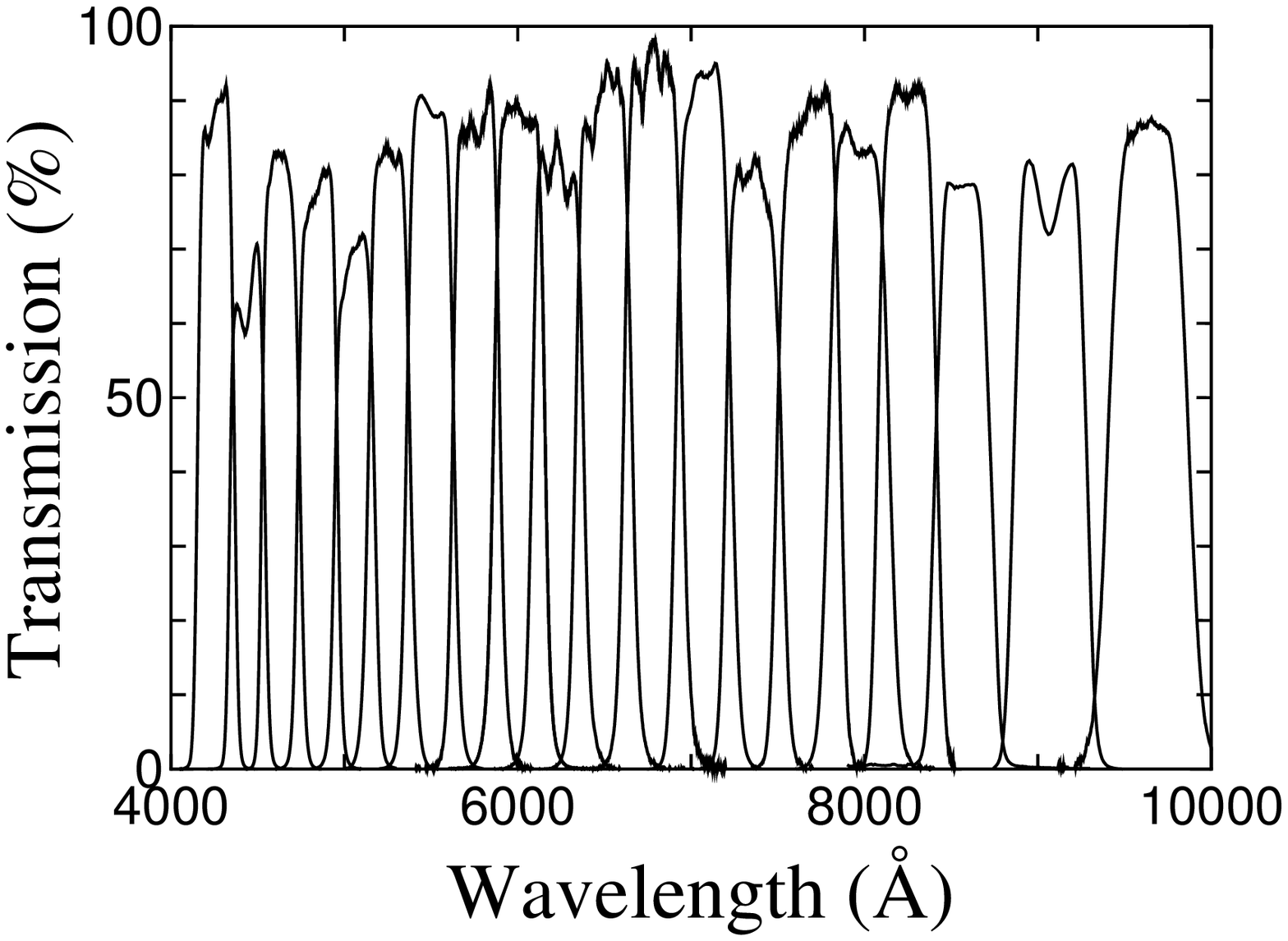}
 \end{center}

 \caption{The filter response curves themselves measured at the center
 of the 20 intermediate-band filters.  Note that these curves do not
 include the effects of the CCD sensitivity, the atmospheric
 transmission, and the transmission of the telescope and the
 instrument.\label{filter_response_all}}

\end{figure*}

 The IA filter system consists of 20 intermediate band filters with a
 spectral resolution of $R = 20$--26, covering 410~nm to
 1000~nm. (Table~\ref{IAfilters}).  The filter response curves are
 shown in Figure~\ref{filter_response_all}.  Note that these response
 curves are those of the filters themselves; that is, the effects of
 the CCD sensitivity, the atmospheric transmission, and the
 transmission of the telescope and the instrument are not included.

%-----------------------------------------------------
%    Table A1
%-----------------------------------------------------
\begin{table}
 \centering
 \renewcommand{\thetable}{A1}
 \tbl{A summary of IA filters.}{
 \begin{tabular*}{45mm}{@{\extracolsep{\fill}}ccc}
  \hline
  Band & $\lambda_{\rm c}^a$ & FWHM$^b$  \\
  & (\AA) & (\AA) \\
  \hline
  $IA427$ & 4271 & 210 \\
  $IA445$ & 4456 & 203 \\
  $IA464$ & 4636 & 217 \\
  $IA484$ & 4842 & 227 \\
  $IA505$ & 5063 & 232 \\
  $IA527$ & 5272 & 242 \\
  $IA550$ & 5512 & 273 \\
  $IA574$ & 5743 & 271 \\
  $IA598$ & 6000 & 294 \\
  $IA624$ & 6226 & 299 \\
  $IA651$ & 6502 & 322 \\
  $IA679$ & 6788 & 336 \\
  $IA709$ & 7082 & 318 \\
  $IA738$ & 7371 & 322 \\
  $IA767$ & 7690 & 364 \\
  $IA797$ & 7981 & 353 \\
  $IA827$ & 8275 & 340 \\
  $IA856$ & 8566 & 325 \\
  $IA907$ & 9068 & 423 \\
  $IA965$ & 9651 & 469 \\
  \hline
 \end{tabular*}}\label{IAfilters}
  \begin{tabnote}

   $^a$Center wavelength defined as the center of the two wavelengths
   at which the filter transmission becomes the half maximum.

   $^b$FWHM calculated from the same filter response curve used to
   evaluate the center wavelength.
  \end{tabnote}
\end{table}

 All the IA filters were manufactured by Barr Associates Co. Ltd (now,
 Materion Co. Ltd).  The specifications for the IA filters are
 summarized in Table~\ref{Specifications}.  Although some of the
 specifications were found not to be fully satisfied, all the filters
 are highly useful for scientific observations (e.g., Fujita et
 al. 2003; Ajiki et al. 2004; Shioya et al. 2005; Yamada et al. 2005;
 Nagao et al. 2008).  Details of measurements of the filter
 transmission is given in Hayashino et al. (2003).  The measured data
 are available at http://www.awa.tohoku.ac.jp/astro/filter.html.
%-------------------------------------------------------------------------
%              Table A2
%-------------------------------------------------------------------------
\begin{table}
 \renewcommand{\thetable}{A2}
 \tbl{The Specifications for the Subaru IA Filter System.}{
 \begin{tabular}{lll}
  \hline
  Item & \multicolumn{2}{c}{Specification}\\
  \hline
  Clear aperture                    & \multicolumn{2}{l}{$185~\mathrm{mm} \times 150~\mathrm{mm}$}  \\
      Peak transmittance ($T_{\rm peak}$) & \multicolumn{2}{l}{$> 70\%$ ($> 80\%$~goal)} \\
  Homogeneity of $T_{\rm peak}$       & \multicolumn{2}{l}{$< 5\%$}                \\
  Ripple (valley/peak)              & \multicolumn{2}{l}{$> 85\%$}                \\
  Linear change (valley/peak)       & \multicolumn{2}{l}{$> 90\%$}             \\
  $\lambda_\mathrm{eff}$ tolerance   & \multicolumn{2}{l}{$< \pm 0.25\%$ of $\lambda_\mathrm{eff}$}         \\
  FWHM tolerance                    & \multicolumn{2}{l}{$< \pm 0.25\%$ of $\lambda_\mathrm{eff}$}         \\
  Bubble & $d < 0.1~\mathrm{mm}$        & acceptable              \\
  & $d = 0.1$--$0.2~\mathrm{mm}$ & $\leq 5$~bubbles        \\
  & $d = 0.2$--$0.5~\mathrm{mm}$ & $\leq 3$~bubbles        \\
  & $d > 0.5~\mathrm{mm}$        & Not allowed             \\
  Stain              &      \multicolumn{2}{l}{Not allowed}             \\
  \hline
  \end{tabular}} \label{Specifications}
\end{table}
%------------------------------------------------------------------
%----Fig A1

%\newpage
\renewcommand{\thesection}{B}
\section{IA-band magnitudes of the standard stars}\label{sec:app-B}

In Table~\ref{tab:standards}, we summarize the theoretical IA-band
magnitudes of the standard stars computed from the {\tt CALSPEC}
spectra and the complete system response, in AB magnitudes.  The
uncertainty given here assumes a perfect knowledge of the system
response and are based solely on the {\tt CALSPEC} statistical and
systematic uncertainties.
%-------------------------------------------------------------------------
%              Table B1
%-------------------------------------------------------------------------
\begin{table*}[htbp]
 \centering
 \renewcommand{\thetable}{B1}
 \tbl{Theoretical IA-band magnitudes of the standard stars.}{
 \begin{tabular}{cccccc}
  \hline
  Band & GD~50 & GD~108 & HZ~4 & HZ~21 & HZ~44\\
%\begin{tabular}{lccccc}
%\hline
%\hline
% Bands & GD~50 & GD~108 & HZ~4 & HZ~21 & HZ~44 \\
%\hline
  \hline
$IA427$ & $ 13.690 \pm  0.002 $ & $13.187 \pm  0.002 $ & $14.445 \pm  0.002 $ & $14.191 \pm  0.002 $ & $11.223 \pm  0.001 $ \\
$IA464$ & $ 13.749 \pm  0.002 $ & $13.297 \pm  0.002 $ & $14.252 \pm  0.002 $ & $14.405 \pm  0.002 $ & $11.374 \pm  0.002 $ \\
$IA484$ & $ 13.869 \pm  0.003 $ & $13.414 \pm  0.003 $ & $14.576 \pm  0.004 $ & $14.474 \pm  0.003 $ & $11.447 \pm  0.002 $ \\
$IA505$ & $ 13.893 \pm  0.003 $ & $13.442 \pm  0.003 $ & $14.385 \pm  0.026 $ & $14.528 \pm  0.026 $ & $11.516 \pm  0.025 $ \\
$IA527$ & $ 13.975 \pm  0.003 $ & $13.504 \pm  0.003 $ & $14.394 \pm  0.046 $ & $14.605 \pm  0.047 $ & $11.583 \pm  0.045 $ \\
$IA574$ & $ 14.164 \pm  0.003 $ & $13.650 \pm  0.003 $ & $14.534 \pm  0.001 $ & $14.771 \pm  0.001 $ & $11.747 \pm  0.001 $ \\
$IA624$ & $ 14.309 \pm  0.002 $ & $13.774 \pm  0.002 $ & $14.662 \pm  0.001 $ & $14.919 \pm  0.001 $ & $11.896 \pm  0.001 $ \\
$IA679$ & $ 14.485 \pm  0.003 $ & $13.917 \pm  0.003 $ & $14.833 \pm  0.001 $ & $15.097 \pm  0.001 $ & $12.073 \pm  0.001 $ \\
$IA709$ & $ 14.559 \pm  0.003 $ & $13.984 \pm  0.003 $ & $14.849 \pm  0.001 $ & $15.176 \pm  0.001 $ & $12.154 \pm  0.001 $ \\
$IA738$ & $ 14.634 \pm  0.002 $ & $14.052 \pm  0.002 $ & $14.912 \pm  0.001 $ & $15.251 \pm  0.001 $ & $12.231 \pm  0.001 $ \\
$IA767$ & $ 14.698 \pm  0.003 $ & $14.119 \pm  0.003 $ & $14.980 \pm  0.001 $ & $15.339 \pm  0.001 $ & $12.321 \pm  0.001 $ \\
$IA827$ & $ 14.844 \pm  0.002 $ & $14.237 \pm  0.002 $ & $15.093 \pm  0.002 $ & $15.491 \pm  0.002 $ & $12.468 \pm  0.001 $ \\
$NB711$ & $ 14.566 \pm  0.004 $ & $13.992 \pm  0.004 $ & $14.860 \pm  0.002 $ & $15.194 \pm  0.003 $ & $12.167 \pm  0.002 $ \\
  \hline
 \end{tabular}}\label{tab:standards}
\end{table*}


\begin{thebibliography}{}


\bibitem[Ajiki et al.(2004)]{2004PASJ...56..597A}
		Ajiki, M., Taniguchi, Y.,
		Fujita, S.~S., et al.\ 2004, \pasj, 56, 597

\bibitem[Amor{\'{\i}}n et al.(2014)]{2014ApJ...788L...4A}
		Amor{\'{\i}}n, R., Grazian, A., Castellano, M., et
		al.\ 2014, \apjl, 788, L4

\bibitem[Amor{\'{\i}}n et al.(2015)]{2015A&A...578A.105A}
		Amor{\'{\i}}n, R., P{\'e}rez-Montero, E., Contini, T.,
		et al.\ 2015, \aap, 578, A105

\bibitem[Atek et al.(2011)]{2011ApJ...743..121A} Atek, H., Siana,
		B.,Scarlata, C., et al.\ 2011, \apj, 743, 121

\bibitem[Bertin \& Arnouts(1996)]{1996A&AS..117..393B} Bertin, E., \&
		Arnouts, S.\ 1996, \aaps, 117, 393

\bibitem[Bohlin(1996)]{1996AJ....111.1743B} Bohlin, R.~C.\ 1996, \aj,
		111, 1743

\bibitem[Bohlin et al.(2001)]{2001AJ....122.2118B} Bohlin, R.~C.,
		Dickinson, M.~E., \& Calzetti, D.\ 2001, \aj, 122,
		2118

\bibitem[Capak et al.(2007)]{2007ApJS..172...99C} Capak, P., Aussel,
		H., Ajiki, M., et al.\ 2007, \apjs, 172, 99

\bibitem[Cardamone et al.(2009)]{2009MNRAS.399.1191C} Cardamone, C.,
		Schawinski, K., Sarzi, M., et al.\ 2009, \mnras, 399,
		1191

\bibitem[Cardamone et al.(2010)]{2010ApJS..189..270C} Cardamone,
		C.~N., van Dokkum, P.~G., Urry, C.~M., et al.\ 2010,
		\apjs, 189, 270

\bibitem[Elvis et al.(2009)]{2009ApJS..184..158E} Elvis, M., Civano,
		F., Vignali, C., et al.\ 2009, \apjs, 184, 158

\bibitem[Feruglio et al.(2010)]{2010ApJ...721..607F} Feruglio, C.,
		Aussel, H., Le Floc'h, E., et al.\ 2010, \apj, 721,
		607

\bibitem[Fujita et al.(2003)]{2003AJ....125...13F} Fujita, S.~S.,
		Ajiki, M., Shioya, Y., et al.\ 2003, \aj, 125, 13

\bibitem[Georgakakis et al.(2014)]{2014MNRAS.443.3327G} Georgakakis,
		A., Mountrichas, G., Salvato, M., et al.\ 2014,
		\mnras, 443, 3327

\bibitem[Hasinger et al.(2007)]{2007ApJS..172...29H} Hasinger, G.,
		Cappelluti, N., Brunner, H., et al.\ 2007, \apjs, 172,
		29

\bibitem[Hayashino et al.(2003)]{2003PNAOJ...7...33H} Hayashino, T.,
		Tamura, H., Matsuda, Y., et al.\ 2003, Publications of
		the National Astronomical Observatory of Japan, 7, 33

\bibitem[Hayashino et al.(2000)]{2000SPIE.4008..397H} Hayashino, T.,
		Taniguchi, Y., Yamada, T., et al.\ 2000, \procspie,
		4008, 397

\bibitem[Hogg et al.(1998)]{1998AJ....115.1418H} Hogg, D.~W., Cohen,
		J.~G., Blandford, R., et al.\ 1998, \aj, 115, 1418

\bibitem[Hopkins(2004)]{2004ApJ...615..209H} Hopkins, A.~M.\ 2004,
		\apj, 615, 209

\bibitem[Hu et al.(2010)]{2010ApJ...725..394H} Hu, E.~M., Cowie,
		L.~L., Barger, A.~J., et al.\ 2010, \apj, 725, 394

\bibitem[Hu et al.(2004)]{2004AJ....127..563H} Hu, E.~M., Cowie,
		L.~L., Capak, P., et al.\ 2004, \aj, 127, 563

\bibitem[Ideue et al.(2009)]{2009ApJ...700..971I} Ideue, Y., Nagao,
		T., Taniguchi, Y., et al.\ 2009, \apj, 700, 971

\bibitem[Ideue et al.(2012)]{2012ApJ...747...42I} Ideue, Y.,
		Taniguchi, Y., Nagao, T., et al.\ 2012, \apj, 747, 42

\bibitem[Iye et al.(2004)]{2004PASJ...56..381I} Iye, M., Karoji, H.,
		Ando, H., et al.\ 2004, \pasj, 56, 381

\bibitem[Ilbert et al.(2006)]{2006A&A...457..841I} Ilbert, O.,
		Arnouts, S., McCracken, H.~J., et al.\ 2006, \aap,
		457, 841

\bibitem[Ilbert et al.(2009)]{2009ApJ...690.1236I} Ilbert, O., Capak,
		P., Salvato, M., et al.\ 2009, \apj, 690, 1236

%\bibitem[Ilbert et al.(2013)]{2013A&A...556A..55I} Ilbert, O.,
%		McCracken, H.~J., Le F{\`e}vre, O., et al.\ 2013,
%		\aap, 556, A55

%\bibitem[Ilbert et al.(2010)]{2010ApJ...709..644I} Ilbert, O.,
%Salvato, M., Le Floc'h, E., et al.\ 2010, \apj, 709, 644

\bibitem[Kaifu et al.(2000)]{2000PASJ...52....1K} Kaifu, N., Usuda,
		T., Hayashi, S.~S., et al.\ 2000, \pasj, 52, 1


\bibitem[Kajisawa et al.(2013)]{2013ApJ...768...51K} Kajisawa, M.,
		Shioya, Y., Aida, Y., et al.\ 2013, \apj, 768, 51

\bibitem[Kakazu et al.(2007)]{2007ApJ...668..853K} Kakazu, Y., Cowie,
		L.~L., \& Hu, E.~M.\ 2007, \apj, 668, 853

\bibitem[Kashikawa et al.(2004)]{2004PASJ...56.1011K} Kashikawa, N.,
		Shimasaku, K., Yasuda, N., et al.\ 2004, \pasj, 56,
		1011

\bibitem[Kobayashi et al.(2015)]{2015ApJ...submitted} Kobayashi,
		M. A. R., Murata, K. L., Koekemoer, A. M., et al.\
		2015, submitted to \apj

\bibitem[Kodaira et al.(2003)]{2003PASJ...55L..17K} Kodaira, K.,
		Taniguchi, Y., Kashikawa, N., et al.\ 2003, \pasj, 55,
		L17

\bibitem[Koekemoer et al.(2007)]{2007ApJS..172..196K} Koekemoer,
		A.~M., Aussel, H., Calzetti, D., et al.\ 2007, \apjs,
		172, 196

\bibitem[Kova{\v c} et al.(2014)]{2014MNRAS.438..717K} Kova{\v c}, K.,
		Lilly, S.~J., Knobel, C., et al.\ 2014, \mnras, 438,
		717

\bibitem[McCracken et al.(2012)]{2012A&A...544A.156M} McCracken,
		H.~J., Milvang-Jensen, B., Dunlop, J., et al.\ 2012,
		\aap, 544, A156

\bibitem[Miyazaki et al.(2012)]{2012SPIE.8446E..0ZM} Miyazaki, S.,
		Komiyama, Y., Nakaya, H., et al.\ 2012, \procspie,
		8446,

\bibitem[Miyazaki et al.(2002)]{2002PASJ...54..833M} Miyazaki, S.,
		Komiyama, Y., Sekiguchi, M., et al.\ 2002, \pasj, 54,
		833

\bibitem[Mobasher et al.(2007)]{2007ApJS..172..117M} Mobasher, B.,
		Capak, P., Scoville, N.~Z., et al.\ 2007, \apjs, 172,
		117

\bibitem[Moles et al.(2008)]{2008AJ....136.1325M} Moles, M.,
		Ben{\'{\i}}tez, N., Aguerri, J.~A.~L., et al.\ 2008,
		\aj, 136, 1325

\bibitem[Molino et al.(2014)]{2014MNRAS.441.2891M} Molino, A.,
		Ben{\'{\i}}tez, N., Moles, M., et al.\ 2014, \mnras,
		441, 2891

\bibitem[Monet et al.(2003)]{2003AJ....125..984M} Monet, D.~G.,
		Levine, S.~E., Canzian, B., et al.\ 2003, \aj, 125,
		984

\bibitem[Murata et al.(2014)]{2014ApJ...786...15M} Murata, K.~L.,
		Kajisawa, M., Taniguchi, Y., et al.\ 2014, \apj, 786,
		15

\bibitem[Murayama et al.(2007)]{2007ApJS..172..523M} Murayama, T.,
		Taniguchi, Y., Scoville, N.~Z., et al.\ 2007, \apjs,
		172, 523

\bibitem[Nagao et al.(2007)]{2007A&A...468..877N} Nagao, T., Murayama,
		T., Maiolino, R., et al.\ 2007, \aap, 468, 877

\bibitem[Nagao et al.(2008)]{2008ApJ...680..100N} Nagao, T., Sasaki,
		S.~S., Maiolino, R., et al.\ 2008, \apj, 680, 100

%\bibitem[Radovich(2002)]{radovich02} Radovich, M. 2002, Astrometrix,
%		Tech. rep., Osservatorio Astronomico di Capodimonte

\bibitem[Radovich et al.(2001)]{2001ASPC..232..297R} Radovich, M., 
Bonnarel, F., Mellier, Y., et al.\ 2001, The New Era of Wide Field 
Astronomy, 232, 297 


\bibitem[Salvato et al.(2009)]{2009ApJ...690.1250S} Salvato, M.,
		Hasinger, G., Ilbert, O., et al.\ 2009, \apj, 690,
		1250

\bibitem[Salvato et al.(2011)]{2011ApJ...742...61S} Salvato, M.,
		Ilbert, O., Hasinger, G., et al.\ 2011, \apj, 742, 61

\bibitem[Sanders et al.(2007)]{2007ApJS..172...86S} Sanders, D.~B.,
		Salvato, M., Aussel, H., et al.\ 2007, \apjs, 172, 86

\bibitem[Schinnerer et al.(2004)]{2004AJ....128.1974S} Schinnerer, E.,
		Carilli, C.~L., Scoville, N.~Z., et al.\ 2004, \aj, 128,
		1974

\bibitem[Schinnerer et al.(2007)]{2007ApJS..172...46S} Schinnerer, E.,
		Smol{\v c}i{\'c}, V., Carilli, C.~L., et al.\ 2007,
		\apjs, 172, 46

\bibitem[Schlegel et al.(1998)]{1998ApJ...500..525S} Schlegel, D.~J.,
                Finkbeiner, D.~P., \& Davis, M.\ 1998, \apj, 500, 525

\bibitem[Scoville et al.(2007)]{2007ApJS..172...38S} Scoville, N.,
		Abraham, R.~G., Aussel, H., et al.\
		2007a, \apjs, 172, 38

\bibitem[Scoville et al.(2013)]{2013ApJS..206....3S} Scoville, N.,
		Arnouts, S., Aussel, H., et al.\ 2013, \apjs, 206, 3

\bibitem[Scoville et al.(2007)]{2007ApJS..172....1S} Scoville, N.,
		Aussel, H., Brusa, M., et al.\
		2007b, \apjs, 172, 1

%\bibitem[Scoville \& Murchikova(2013)]{2013ApJ...779...75S} Scoville,
%		N., \& Murchikova, L.\ 2013, \apj, 779, 75

\bibitem[Shimasaku et al.(2005)]{2005PASJ...57..447S} Shimasaku, K.,
		Ouchi, M., Furusawa, H., et al.\ 2005, \pasj, 57, 447

\bibitem[Shioya et al.(2005)]{2005PASJ...57..287S} Shioya, Y.,
		Taniguchi, Y., Ajiki, M., et al.\ 2005, \pasj, 57, 287

\bibitem[Shioya et al.(2008)]{2008ApJS..175..128S} Shioya, Y.,
		Taniguchi, Y., Sasaki, S.~S., et al.\ 2008, \apjs,
		175, 128

\bibitem[Shioya et al.(2009)]{2009ApJ...696..546S} Shioya, Y.,
		Taniguchi, Y., Sasaki, S.~S., et al.\ 2009, \apj, 696,
		546

\bibitem[Smol{\v c}i{\'c} et al.(2012)]{2012A&A...548A...4S} Smol{\v
		c}i{\'c}, V., Aravena, M., Navarrete, F., et al.\
		2012, \aap, 548, A4

\bibitem[Takahashi et al.(2007)]{2007ApJS..172..456T} Takahashi,
		M.~I., Shioya, Y., Taniguchi, Y., et al.\ 2007, \apjs,
		172, 456

\bibitem[Taniguchi(2004)]{2004sgyu.conf..107T} Taniguchi, Y.\ 2004,
		Studies of Galaxies in the Young Universe with New
		Generation Telescope, Proceedings of Japan-German
		Seminar, held in Sendai, Japan, July 24-28, 2001,
		Eds.: N. Arimoto and W. Duschl, 2004, p. 107-111

\bibitem[Taniguchi et al.(2005)]{2005PASJ...57..165T} Taniguchi, Y.,
		Ajiki, M., Nagao, T., et al.\ 2005, \pasj, 57, 165


\bibitem[Taniguchi et al.(2015)]{2015ApJ...809L...7T} Taniguchi, Y.,
		Kajisawa, M., Kobayashi, M.~A.~R., et al.\ 2015,
		\apjl, 809, L7


\bibitem[Taniguchi et al.(2009)]{2009ApJ...701..915T} Taniguchi, Y.,
		Murayama, T., Scoville, N.~Z., et al.\ 2009, \apj,
		701, 915

\bibitem[Taniguchi et al.(2007)]{2007ApJS..172....9T} Taniguchi, Y.,
		Scoville, N., Murayama, T., et al.\ 2007, \apjs, 172,
		9 (Paper~I)

\bibitem[Taniguchi et al.(2010)]{2010ApJ...724.1480T} Taniguchi, Y.,
		Shioya, Y., \& Trump, J.~R.\ 2010, \apj, 724, 1480

\bibitem[Wolf et al.(2004)]{2004A&A...421..913W} Wolf, C.,
		Meisenheimer, K., Kleinheinrich, M., et al.\ 2004,
		\aap, 421, 913

\bibitem[Wolf et al.(2003)]{2003A&A...408..499W} Wolf, C., Wisotzki,
		L., Borch, A., et al.\ 2003, \aap, 408, 499

\bibitem[Yamada et al.(2005)]{2005PASJ...57..881Y} Yamada, S.~F.,
		Sasaki, S.~S., Sumiya, R., et al.\ 2005, \pasj, 57,
		881

\bibitem[Zamojski et al.(2007)]{2007ApJS..172..468Z} Zamojski, M.~A.,
		Schiminovich, D., Rich, R.~M., et al.\ 2007, \apjs,
		172, 468

\end{thebibliography}
\end{document}